\documentclass[useAMS,usenatbib]{mn2e}
\usepackage{natbib}
\usepackage{graphicx}
\usepackage{amsmath}
\usepackage[T1]{fontenc}
\DeclareGraphicsExtensions{.pdf}

\title[21 cm emission - LAE cross-power spectrum]{Predictions for the 21cm-galaxy cross-power spectrum observable with SKA and future galaxy surveys}
\author[D. Vrbanec et al.]{Dijana Vrbanec$^{1}$,\thanks{E-mail:
dvrbanec@mpa-garching.mpg.de} Benedetta Ciardi$^{1}$, Vibor Jeli\'{c}$^{2}$,
Hannes Jensen$^3$, Ilian T. Iliev$^4$, \newauthor Garrelt Mellema$^3$, Saleem Zaroubi$^{5,6,7}$\\
$^1$Max Planck Institute for Astrophysics, Karl-Schwarzschild-Strasse 1, D-85748 Garching bei M\"{u}nchen, Germany\\
$^2$Ru{\dj}er Bo\v{s}kovi\'{c} Institute, Bijeni\v{c}ka cesta 54, 10000 Zagreb, Croatia \\
$^3$Department of Astronomy and Oskar Klein Centre, Stockholm University, AlbaNova, SE-10691 Stockholm, Sweden\\
$^4$Astronomy Centre, Department of Physics and Astronomy, Pevensey II Building, University of Sussex, Falmer, Brighton BNI 9QH, UK\\
$^5$Kapteyn Astronomical Institute, University of Groningen, PO Box 800, NL-9700 AV Groningen, the Netherlands\\
$^6$Department of Natural Sciences, Open University of Israel, 1 University Road, PO Box 808, Ra'anana 4353701, Israel\\
$^7$Department of Physics, The Technion, Haifa 32000, Israel\\
}
\begin{document}

\date{Accepted - Received -; in original form -}

\pagerange{\pageref{firstpage}--\pageref{lastpage}} \pubyear{2014}

\maketitle

\label{firstpage}

\begin{abstract}
In this paper we use radiative transfer + N-body simulations to explore the feasibility of measurements of cross-correlations between the 21~cm field observed by the Square Kilometer Array (SKA) and high-$z$ Lyman Alpha Emitters (LAEs) detected in galaxy surveys with the Subaru Hyper Supreme Cam (HSC), Subaru Prime Focus Spectrograph (PFS) and Wide Field Infrared Survey Telescope (WFIRST). 21cm-LAE cross-correlations are in fact a powerful probe of the epoch of reionization as they are expected to provide precious information on the progress of reionization and the typical scale of ionized regions at different redshifts.
The next generation observations with SKA will have a noise level much lower than those with its precursor radio facilities, introducing a significant improvement in the measurement of the cross-correlations. We find that an SKA-HSC/PFS observation will allow to investigate scales below $\sim 10\ h^{-1}$~Mpc and $\sim 60\ h^{-1}$~Mpc at $z=7.3$ and 6.6, respectively.  
WFIRST will allow to access also higher redshifts, as it is expected to observe spectroscopically $\sim$900 LAEs per square degree and unit redshift in the range $7.5\le z\le 8.5$. Because of the reduction of the shot noise compared to HSC and PFS, observations with WFIRST will result in more precise cross-correlations and increased observable scales. 
\end{abstract}


\section{Introduction}

The detection of the 21~cm line from high-redshift neutral hydrogen (HI) is one of the last observational frontiers and the goal of several present and upcoming radio facilities such as the LOw Frequency ARray\footnote{www.lofar.org} (LOFAR), the Murchison Widefield Array\footnote{http://www.mwatelescope.org} (MWA),  the Hydrogen Epoch of Reionization Array\footnote{https://reionization.org} (HERA), and the Square Kilometer Array\footnote{https://www.skatelescope.org} (SKA).
Such observational campaigns will provide the first constraints on the history of hydrogen reionization and the evolution of the properties of the high-$z$ intergalactic medium (IGM) (see \citealt{Pritchard2012} for a review). 
Tomography of the 21~cm line will offer information on the temperature and ionization state of the IGM, on the topology of reionization and possibly on the properties of the sources of ionizing photons (see e.g. \citealt{Tozzi.Madau.Meiksin.Rees_2000,Ciardi.Madau_2003,Zaroubi_etal_2012}), while the detection (or even an upper limit) of fluctuations and power spectrum of the differential brightness temperature will provide invaluable insight on the timing of reionization and  statistical estimates of its properties (e.g. \citealt{Madau1997,Mellema.Iliev.Pen.Shapiro_2006,Baek_etal_2009,Patil_etal_2014}). The amount of HI along lines of sight towards high-$z$ radio loud sources could instead be measured with 21~cm absorption systems (e.g. \citealt{Carilli2002,Furlanetto_2006,Xu_etal_2009,Ciardi_etal_2013,Semelin_2015}).

It has also been suggested that correlating the 21~cm signal with observations in different frequency bands would reduce systematic effects and confirm the cosmological origin of the signal, in addition to potentially provide additional and independent information on both the reionization process and the sources it is being correlated with. Investigations have been made with respect to correlations with the cosmic microwave background radiation \citep{Salvaterra.Ciardi.Ferrara.Baccigalupi_2005,Jelic2010,Ma_etal_2018}, high-$z$ galaxies \citep{Lidz2009,W2013,Vrbanec2016,SobacchiMesingerGreig2016,Hutter.Dayal.Muller.Trott_2017,Kubota_etal_2018}, NIR \citep{Fernandez2014} and X-ray \citep{Ma.Ciardi.Eide.Helgason_2018} background radiation.
Particular attention has been devoted to the possibility of cross-correlating the 21~cm signal with Ly$\alpha$ emitters (LAEs) surveys, as the 21~cm field is expected to anti-correlate (in terms of cross-power spectra and cross-correlation coefficients) with the galaxy field on large scales (where most of the HI lies once reionization is well underway but where there is a paucity of galaxies) and become roughly un-correlated on small scales, within the ionized regions. 
More specifically, \cite{SobacchiMesingerGreig2016} investigated the effect of morphology and the mass of LAEs hosting halos on the cross-correlations, showing that LOFAR or SKA in combination with the Subaru's Hyper Suprime-Cam (HSC) should be able to distinguish between a half-neutral and a completely ionized universe. \cite{Hutter_etal_2018} expanded on that investigation demonstrating that neutral hydrogen fractions of 0.1, 0.25, 0.5 could be distinguished. \cite{Kubota_etal_2018} confirmed the detectability of the cross-correlations with MWA, SKA, HSC and the Subaru Prime Focus Spectrograph (PFS)\footnote{http://pfs.ipmu.jp/}, focusing on the error budget of these instruments and possible LAEs observational strategies.

In \citet[][hereafter V2016]{Vrbanec2016} we had investigated the feasibility of cross-correlations at $z=6.6$ and 7.3 with LOFAR and HSC, as an agreement was in place among the two teams to observe the common field ELAIS-N1. 
As the LOFAR peak performance turned out to be at redshifts higher than initially envisioned (i.e. >$z\sim 8.5-10.5$, \citealt{Patil_etal_2017}), this has compromised the ability of a cross-correlation exercise, discussed in V2016, to yield significant detection.
Therefore, we extend the investigation performed in V2016 to the SKA with the HSC, the PFS and the Wide Field Infrared Survey Telescope (WFIRST)\footnote{https://wfirst.gsfc.nasa.gov/}.

The paper is structured as follows. In section~2 we present the simulations and methodology adopted, in section 3 the results, followed by a discussion in section 4. Our conclusions are summarized in section 5.

\section{Methodology}

In our cross-correlation analysis we follow closely what was done in V2016. We thus refer the reader to that paper for more details.

To compute the cross-correlations we have used the simulation of reionization described in \cite{Iliev2014}. This has been run in a box of length $425 h^{-1}$~cMpc (equivalent to $\sim$ 4 deg at $z$=7) with 165 billion particles distributed on a $10,976^3$ cells grid ($3.9\,h^{-1}$kpc gravity force resolution). 
The N-body simulation has been initialized at $z=300$ using the Zel'dovich approximation together with a power spectrum of the linear fluctuations generated with the code CAMB \citep{Lewis2000}. The outputs of the simulation have been mapped onto a grid of $504^3$ cells and post-processed with the radiative transfer code C$^2$-RAY \citep{Mellema2006a} to model IGM reionization. Finally, the results have been used as input to a Ly$\alpha$ radiative transfer method to study high-$z$ LAEs and their observability  (\citealt{Jensen2013, Jensen2014}). The LAEs so modelled reproduce the observed luminosity functions. The reader should refer to the original papers for more details.

As in V2016, we evaluate the differential brightness temperature associated to the neutral hydrogen distribution obtained from the simulations as (e.g. \citealt{Field1959,Madau1997,Furlanetto2006a}):
\begin{multline}
\delta T_b= 28.5\ \mathrm{mK}\ (1+\delta)x_{HI}\left(\frac{\Omega_b}{0.042}\frac{h}{0.73}\right)\\\times\left[\left(\frac{1+z}{10}\right)\left( \frac{0.24}{\Omega_m} \right)\right]^{1/2},
\label{eq:tb}
\end{multline}
where $x_{HI}(1+\delta)=n_{HI}/\langle n_H\rangle$ is the average density of neutral hydrogen in units of the average density of hydrogen at redshift $z$, and we assume the spin temperature to be much larger than the temperature of the CMB. $\delta T_b$ is then used to produce mock observations with SKA (see section~3).
Subaru and WFIRST mock observations are instead produced using the Ly$\alpha$ intrinsic and transmitted luminosities extracted from the same simulations (see section~3).

Following \cite{Lidz2009}, we decompose the 21cm-galaxy cross-power spectrum at wave number $k=|k|$, $\Delta^2_{21,gal}(k)$, into three terms:
\begin{multline}
\Delta^2_{21,gal}(k)=\tilde{\Delta}^2_{21,gal}(k)/\delta T_{b0} \\  = \langle x_{HI} \rangle [ \Delta^2_{x_{HI},gal}(k)+\Delta^2_{\rho,gal}(k)+\Delta^2_{x_{HI}\rho,gal}(k)],
\label{eq:cps} 
\end{multline}
where $\Delta^2_{x_{HI},gal}$, $\Delta^2_{\rho,gal}$ and $\Delta^2_{x_{HI}\rho, gal}$ are, respectively, the neutral fraction-galaxy power spectrum, the density-galaxy power spectrum and neutral density-galaxy power spectrum. $\delta T_{b0}$ is the 21~cm brightness temperature relative to the CMB for neutral gas at the mean density, and $\langle x_{HI} \rangle$ is the volume-averaged HI fraction. Given the fields $a$ and $b$, $\Delta^2_{a,b}$ is their dimensionless cross-power spectrum, and it is written as $\Delta^2_{a,b}(k)=k^3P_{a,b}(k)/{2\pi^2}$ in 3D, and $\Delta^2_{a,b}(k)= k^2 P_{a,b}(k)/\pi$ in 2D. Here $P_{a,b}$ is the dimensional cross-power spectrum of the two fields, which are represented in terms of their fluctuations at any given location $r$, i.e. $\delta_a(r)=(a(r)-\langle a\rangle)/\langle a \rangle$, and the same for $b$\footnote{The theoretical cross-power spectrum is evaluated as $\langle
a \rangle=(\sum_{i=1}^N a_i)/N$, with $N$ number of pixels contained in the part of the simulation considered. With the exception of the galaxy field in mock observations, we calculate all quantities in this way. The galaxy field is evaluated as $\langle N_{gal} \rangle =  N_{gal}/V$, with $V$ volume of the survey and $N_{gal}$ number of galaxies in the mock observation. This choice was made because of an easier comparison to the shot noise of the power spectrum, $P_{shot}(k)=1/n_{gal}$, where $n_{gal}$ is the average number of galaxies in the volume covered by the survey.}. We refer the reader to \citet{Lidz2009} and V2016 for a  more detailed discussion of the various terms.

In addition, we have evaluated the cross-correlation function, which characterizes the changes in real space of the correlation between two fields, $a$ and $b$. We write their cross-correlation function as
$\xi_{a,b}(\mathbf{r})=\langle\delta_a(\mathbf{x})\delta_b(\mathbf{x}+\mathbf{r})\rangle$,
where $\delta_a(\mathbf{x})$ and $\delta_b(\mathbf{x})$ are the fractional fluctuation of field $a$ and $b$ at location $\mathbf{x}$, respectively.
The observed cross-correlation function is evaluated as:
\begin{equation}
\xi_{21,LAE}(r)=\frac{\sum_{\mathbf{x}}\delta_{LAE}(\mathbf{x})\delta_{21}(\mathbf{x}+\mathbf{r})}{N_{pair}(r)},
\end{equation}
where $N_{pair}(r)$ is the number of 21cm-LAE pairs at a separation $r$, while $\delta_{LAE}$ and $\delta_{21}$ are the fractional fluctuations of the LAE and 21cm fields, respectively.

To evaluate the cross-correlations, we used simulation boxes at $z=$6.68, 7.3, and 8.06, when the volume (mass) averaged ionized fractions are $\langle x \rangle=$ 0.93 (0.95), 0.48 (0.58), and 0.21 (0.30), respectively. These redshifts were chosen because HSC has two narrow-band filters observing at $z= 6.6$ and 7.3, while WFIRST will observe in the redshift range $7.5<z<8.5$. 

To compute a 3D cross-power spectrum we merged bins so that $\Delta k>0.02\ h$~Mpc $^{-1}$, which corresponds to the smallest mode resolved by a field of view (FoV) of 16~deg$^2$, i.e. equivalent to the one covered by our simulations. We also avoided correlations in power induced by the window function\footnote{The sphere that is used to evaluate a spherically averaged $P(k)$ for a simulation box of length 425~$h^{-1}$cMpc, should have an equivalent volume, i.e. a radius $R=264$ $h^{-1}$~cMpc. The first zero of a window function for a spherical top-hat is at $dk\cdot R\sim4.5$, meaning that $k$-values at distances shorter than $4.5/R=0.02\ h$~Mpc$^{-1}$ are correlated \citep{Feldman1994, Furlanetto2007,Lidz2009, W2013}.} by having a binning with $\Delta \log k=0.02$.
As the FoV of HSC and PFS  is 7~deg$^2$ and 1.7~deg$^2$ at $z=$6.6 and 7.3, respectively (see section~3), when evaluating cross-power spectra for these instruments we used $\Delta \log k=0.03$ (0.05) and  $\Delta k>0.04$ (0.07) $h$~Mpc $^{-1}$ for $z=6.6$ (7.3).

\section{Correlations observable with SKA and future galaxy surveys}

While the reader can find in V2016 a discussion on the theoretical correlations (including the effect of 2D projection), in
this section we will analyze the observability of 21cm-galaxy cross-correlations with SKA and galaxy surveys planned with the Subaru HSC and PFS, and with WFIRST. 

Once the differential brightness temperature is evaluated within a simulation box using eq.~\ref{eq:tb}, the theoretical 3D 21cm-galaxy cross-power spectrum can be calculated with eq.~\ref{eq:cps} (see section 3.1 of V2016 for an extensive discussion). 
Finally, to produce mock observations, the various instruments characteristics need to be taken into account.

To build mock observations with SKA, we model the instrument characteristics using the OSKAR simulator\footnote{https://bitbucket.org/account/user/oscaremr/projects/OSCAR} for the preliminary configuration of SKA1-LOW presented in \cite{Chapman2015}. Full correlation between all 866 core stations (with a maximum baseline length of 5.29 km) is assumed to produce the images of the point spread function (PSF), which have been generated with the Common Astronomy Software Applications package (CASA, \citealt{Brogan2007}) across a 5 deg FoV. The coordinates of the baselines have been generated for a 12-hour synthesis observation. The noise is then normalized assuming a 1000 hour integration time following the prescription described in e.g. \cite{Thompson2001}.

In the following we will investigate the cross-correlation between SKA mock observations thus obtained and various galaxy surveys.

\subsection{Subaru Hyper Supreme Cam}
\label{SKA:HSC}

\begin{figure}
\centering
\includegraphics[width=84mm]{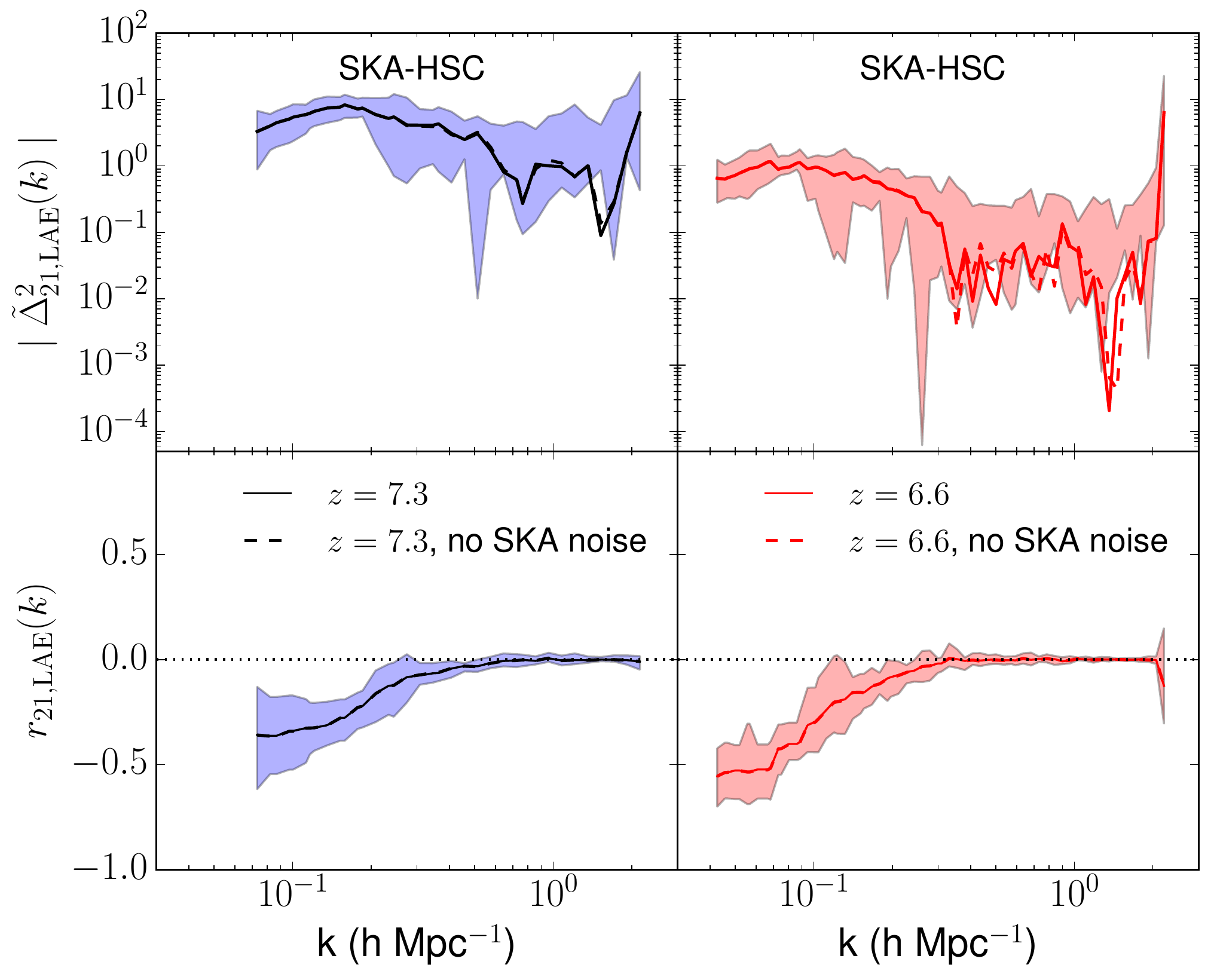}
\caption{{\it Top panels:} 2D unnormalized by $\delta T_{b0}$, circularly averaged 21cm-LAE cross-power spectra between SKA and HSC at $z=7.3$ (left panel) and 6.6 (right). 
{\it Bottom panels:} 21cm-LAE cross-correlation coefficient, $r_{21,\mathrm{LAE}}$, corresponding to $\Delta^2_{21,\mathrm{LAE}}$. 
The field of view is 1.7 deg$^2$ and 7 deg$^2$ at $z=7.3$ and $z=6.6$, respectively. Solid (dashed) lines refer to the cross-power spectrum with (without) SKA noise. Shaded areas indicate scatter from 10 mock observations.
}
\label{SKA:HSCcps}
\end{figure}

In this section we will discuss the observability of the cross-correlation between SKA and HSC.

HSC will observe four fields of 7~deg$^2$ at $z=6.6$ as part of the deep layer, and four fields of 1.7~deg$^2$ (two at $z=6.6$ and two at $z=7.3$) as part of the ultradeep layer. Observations are made with narrow-band filters of $\Delta z = 0.1$, corresponding to about one tenth of our simulation length, and thus the  galaxies will appear as if lying on a single plane. In this case then the observed cross-power spectrum will be a circularly averaged 2D one. Following the procedure outlined and tested in V2016, to match the size of the HSC's FoV the box dimension has been reduced by removing its external cells. We then select 1375 (20) galaxies at $z=6.6$ (7.3) with Ly$\alpha$ equivalent width $>20$ \AA ~and luminosity $>2.5 \times 10^{42}$~erg~s$^{-1}$ to mimic the HSC sensitivity and expected number of observed LAEs per field\footnote{Note that, because of the small number of galaxies expected to be observed at $z=7.3$, the shot noise is dominant and the resulting correlations cannot be considered statistically significant.}. Finally, to match the width of the narrow-band filters we have divided the boxes into 10 sub-boxes of 50 slices each, and every sub-box has been collapsed onto a single plane to mimic the fact that HSC observations will provide 2D galaxy maps. 
Each of the 10 HSC mock observation is correlated to the corresponding SKA mock observation, and the resulting 21cm-LAEs cross-power spectra or cross-correlation functions are averaged to get a sample independent result.

The resulting 2D, unnormalized by $\delta T_{b0}$, circularly averaged 21cm-LAE cross-power spectra are shown in Fig. \ref{SKA:HSCcps}, together with the corresponding cross-correlation coefficients, defined as $r_{21,gal}(k)=P_{21,gal}(k)/[P_{21}(k)P_{gal}(k)]^{1/2}$. The expected dependence of the intensity of the cross-correlation on redshift is maintained, as well as a clear anti-correlation at the largest scales. 
Comparing the resulting cross-power spectra with the ones between LOFAR and HSC (Fig. 4 in \citealt{Vrbanec2016}) it is easy to notice that the spectra observed by SKA and HSC are smoother, more similar to the theoretical ones, and with smaller scatter, so that larger scales (up to $k<0.3\ h$~Mpc$^{-1}$) are now accessible and an anti-correlation can be detected also at $z=7.3$. The 21cm-LAE cross-power spectra with and without the SKA noise show very little difference, and the resulting cross-correlation coefficients have an anti-correlation stronger (i.e. closer to the theoretical value) than the one observed with LOFAR and HSC ($\sim-0.4$ at $z=7.3$ and $\sim-0.6$ at $z=6.6$). 
As discussed in V2016, also with SKA as a reference instrument for 21~cm observations, the anti-correlation at $z=6.6$ is stronger than at $z=7.3$ (differently from the theoretical expectations), as here observations are dominated by shot-noise due to the small number of observed galaxies. Additionally, the smaller field of view at $z=7.3$ covers less of the large-scale modes that would boost the anti-correlation strength at large-scales. It should also be noted that the cross-correlation coefficient is more affected by noise than the power spectrum because it contains the contribution from various power spectra. 

\begin{figure}
\centering
\includegraphics[width=84mm]{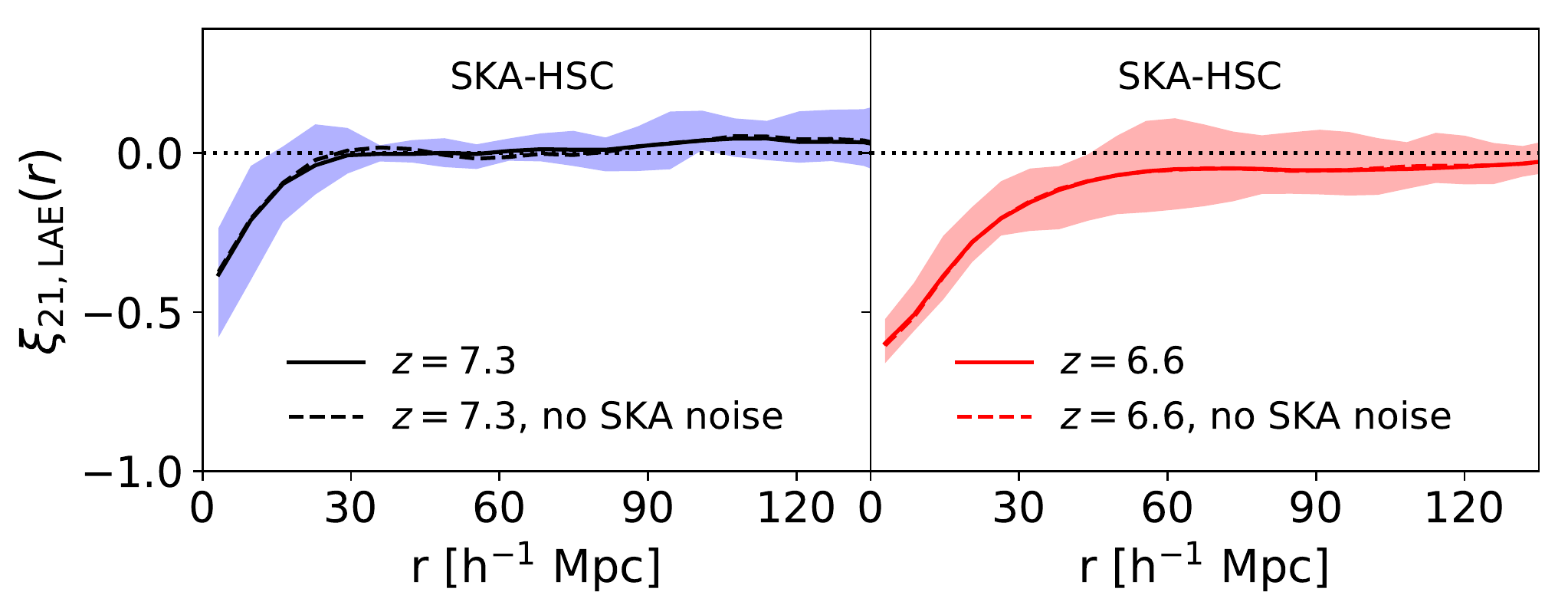}
\caption{2D 21cm-LAE cross-correlation function between SKA and HSC at $z=7.3$ (left panel) and 6.6 (right). Solid (dashed) lines refer to the cross-correlation function with (without) SKA noise. The black dotted lines indicate zero correlation and shaded areas indicate scatter from 10 mock observations.}
\label{SKA:HSCccf}
\end{figure}

In Fig. \ref{SKA:HSCccf} we present the 2D 21cm-LAE cross-correlation functions, which again show a clear anti-correlation on small scales both at $z=6.6$ and 7.3.
Also in this case, a clear improvement is seen in comparison to results expected from LOFAR (see upper panel of Fig.~8 in V2016), for which the scatter estimated from 10 mock observations is large enough to prevent a clear detection at any redshift and scale.

\subsection{Subaru Prime Focus Spectrograph}

\begin{figure}
\centering
\includegraphics[width=84mm]{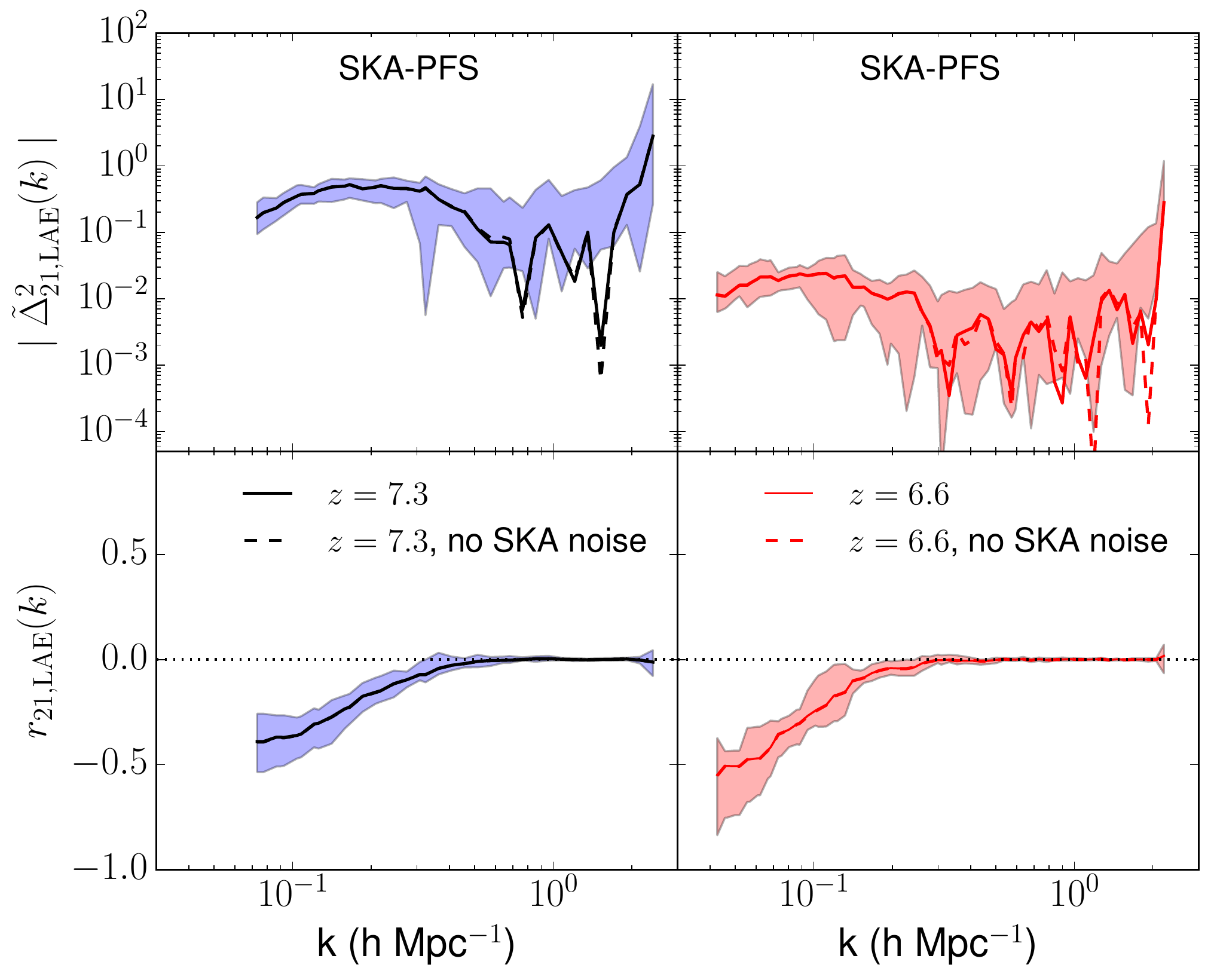}
\caption{{\it Top panels:} 3D unnormalized by $\delta T_{b0}$, spherically averaged 21cm-LAE cross-power spectra between SKA and PFS at $z=7.3$ (left panel) and 6.6 (right). 
{\it Bottom panels:} 21cm-LAE cross-correlation coefficient, $r_{21,\mathrm{LAE}}$, corresponding to $\Delta^2_{21,\mathrm{LAE}}$. 
The field of view is 1.7 deg$^2$ and 7 deg$^2$ at $z=7.3$ and $z=6.6$, respectively. Solid (dashed) lines refer to the cross-power spectrum with (without) SKA noise. Shaded areas indicate scatter from 10 mock observations.}
\label{SKA:PFScps}
\end{figure}

Around 2020 the PFS is expected to be ready for operation. It will focus on the LAEs detected by HSC, but will give the precise position of the objects due to spectroscopic observations. Hence, a 3D instead of 2D cross-correlation can be measured.

As PFS' FoV (7 deg$^2$ at $z=6.6$ and 1.7 deg$^2$ at $z=7.3$), depth ($\Delta z=0.1$), and expected number of observed LAEs per field (1375 at $z=6.6$ and 20 at $z=7.3$) are those of HSC, the 10 sub-boxes obtained from our simulations are the same as those discussed in the previous section, but here the sub-boxes are not collapsed onto a single plane, since PFS will acquire the exact position of the observed LAEs. We thus create 3D mock observations, which are correlated to the corresponding SKA mock observations, resulting in 10 cross-correlations which are then averaged to obtain a sample independent result.

The resulting 3D, unnormalized by $\delta T_{b,0}$, spherically averaged 21cm-LAE cross-power spectra are shown in Fig. \ref{SKA:PFScps}. They look very similar to the ones measured with SKA and HSC (see Fig.~\ref{SKA:HSCcps}). However, the strength of the cross-power spectra is now lower due to the spherical average over a 3D volume, where the reduction of strength is higher at $z=6.6$ due to the larger observed volume. The cross-power spectra become noisy around the same scale as the cross-power spectra measured with SKA and HSC, since they have similar shot noise (due to the same number of observed LAEs; see section 4). The strength of the anti-correlation at large scales in the cross-correlation coefficient is similar to the one measured with SKA and HSC, i.e. $\sim-0.4$ at $z=7.3$ and $\sim-0.6$ at $z=6.6$. 
We should note that the projection effects which enter our calculations when using HSC as reference instrument for LAEs observations do not affect the results. The qualitative behaviour of the various correlations is in fact the same as in the case of PFS, for which 3D rather than 2D power spectra will be available. However, in 3D the scatter is reduced due to projection effects (see Figs. 2 and 3 of V2016). As for HSC, also in this case, the effect of the SKA noise is minimal.

\begin{figure}
\centering
\includegraphics[width=84mm]{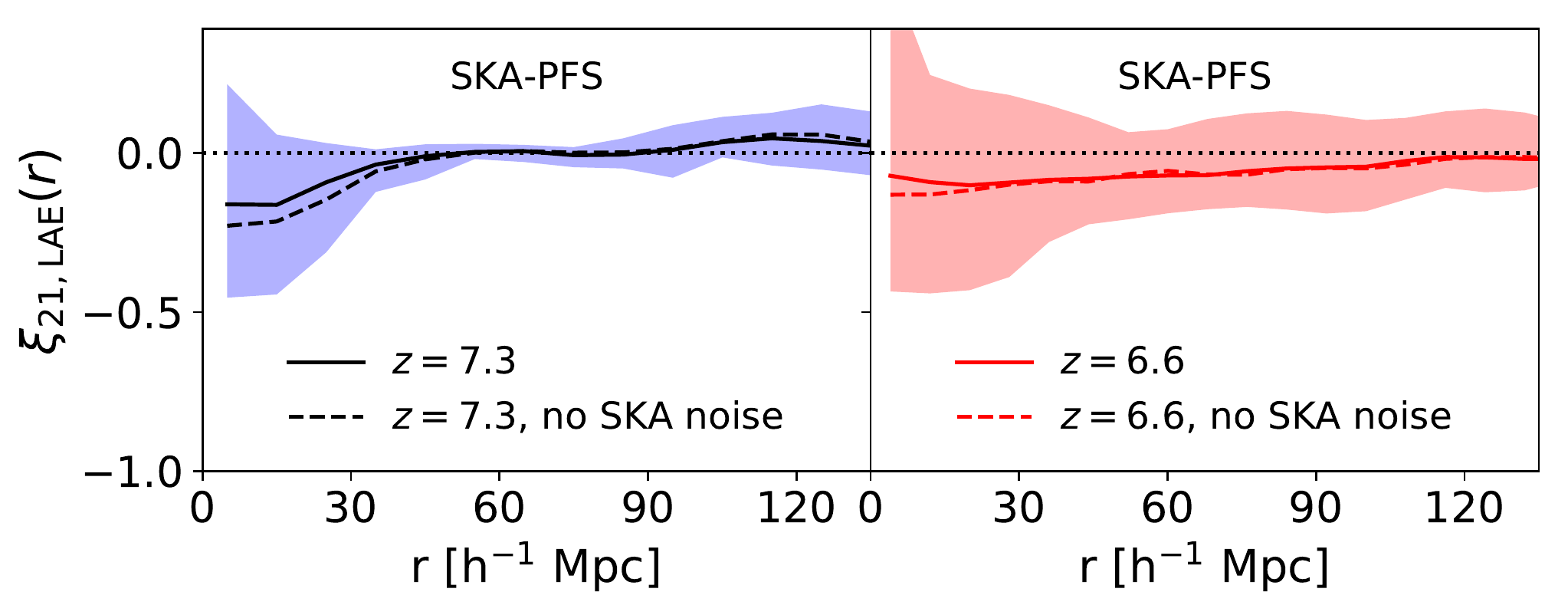}
\caption{3D 21cm-LAE cross-correlation function between SKA and PFS at $z=7.3$ (left panel) and 6.6 (right). Solid (dashed) lines refer to the cross-correlation function with (without) SKA noise. The black dotted lines indicate zero correlation and shaded areas indicate scatter from 10 mock observations.}
\label{SKA:PFSccf}
\end{figure}

In Fig. \ref{SKA:PFSccf} we show the 2D 21cm-LAE cross-correlation functions. Differently from HSC, here the functions are very noisy, due to the higher shot noise (see discussion in Sec.~\ref{sec:discuss}) and do not  offer any valuable information.  

\subsection{Wide Field Infrared Survey Telescope}

The WFIRST is a space facility expected to observe 900 LAEs per square degree and unit redshift in the range $7.5\le z\le 8.5$, with Ly$\alpha$ equivalent width $>$5\AA ~and luminosity $>2.8 \times 10^{42}$~erg~s$^{-1}$. Observations will be done spectroscopically over wide areas of the sky. We constructed a mock observation for WFIRST using a FoV corresponding to our full simulation box, i.e. 16 deg$^2$, a depth of $\Delta z=1$, and 14400 observed LAEs, as expected to be detected within this volume. It should be noted that, differently from the HSC and PFS case, here only one mock observation could be designed because the depth of the experiment corresponds to the entire box.
The SKA mock observation has been created following the usual procedure, using the same FoV and depth of WFIRST, and it has been correlated to the mock WFIRST observation.

\begin{figure}
\centering
\includegraphics[width=84mm]{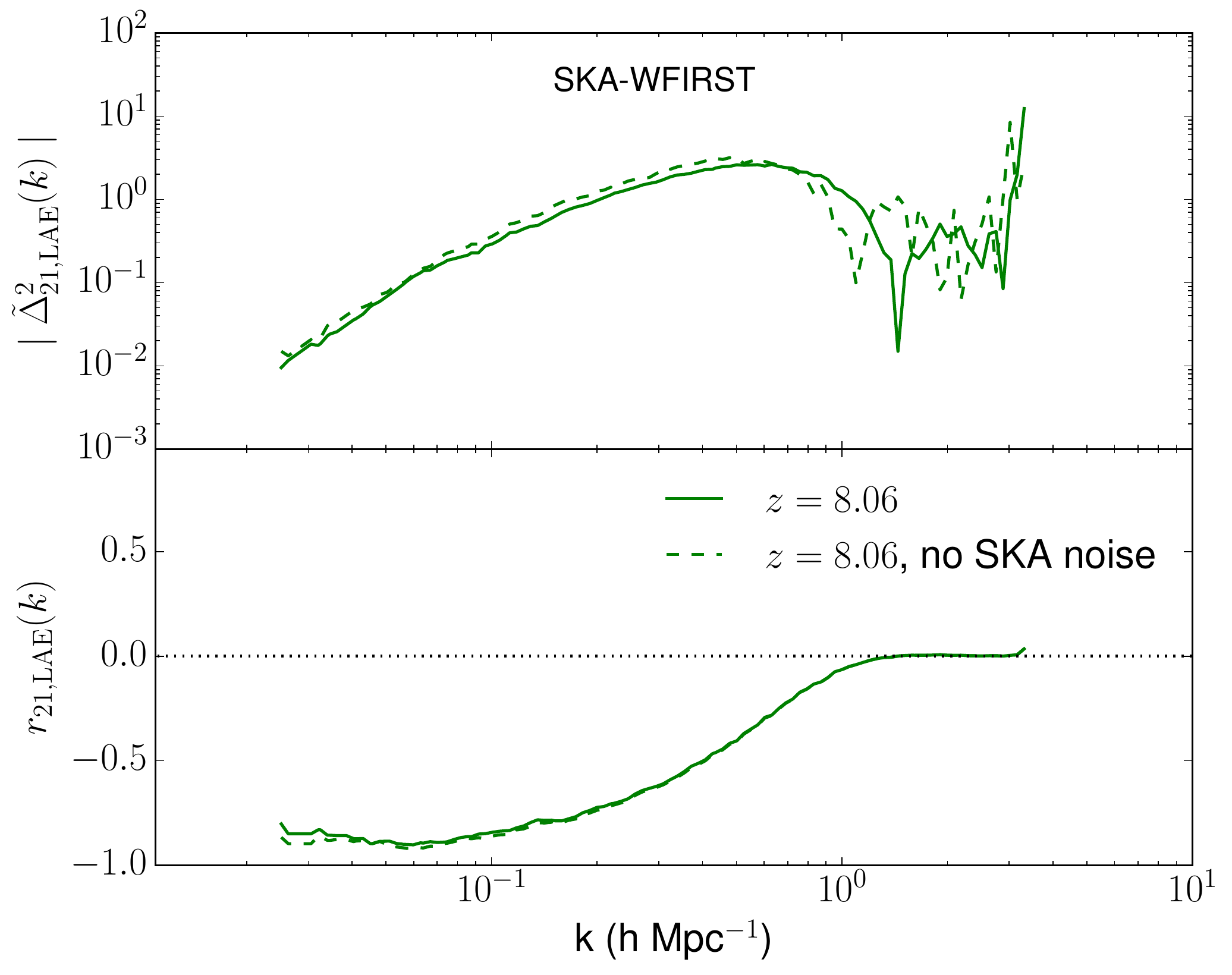}
\caption{{\it Top panel:} 3D unnormalized by $\delta T_{b0}$, spherically averaged 21cm-LAE cross-power spectrum between SKA and WFIRST at $z=8.06$. 
{\it Bottom panel:} 21cm-LAE cross-correlation coefficient, $r_{21,\mathrm{LAE}}$, corresponding to $\Delta^2_{21,\mathrm{LAE}}$. 
The field of view is 16 deg$^2$. The solid (dashed) line refers to the cross-power spectrum with (without) SKA noise.}
\label{SKA:WFIRSTcps}
\end{figure}

The resulting 3D 21cm-LAE cross-power spectrum and cross-correlation coefficient are shown in Fig. \ref{SKA:WFIRSTcps}. Here the 21cm-LAE cross power spectrum has a shape very similar to the theoretical one (Fig. 2 in V2016), although it becomes noisy on small scales, i.e. $k>1\ h$ Mpc$^{-1}$. The corresponding cross-correlation coefficient shows strong anti-correlation on large scales ($r_{21,LAE}\sim-0.9$). As in the previous cases, the SKA noise has very little influence on the resulting cross-power spectrum and cross-correlation coefficient. 

\begin{figure}
\centering
\includegraphics[width=84mm]{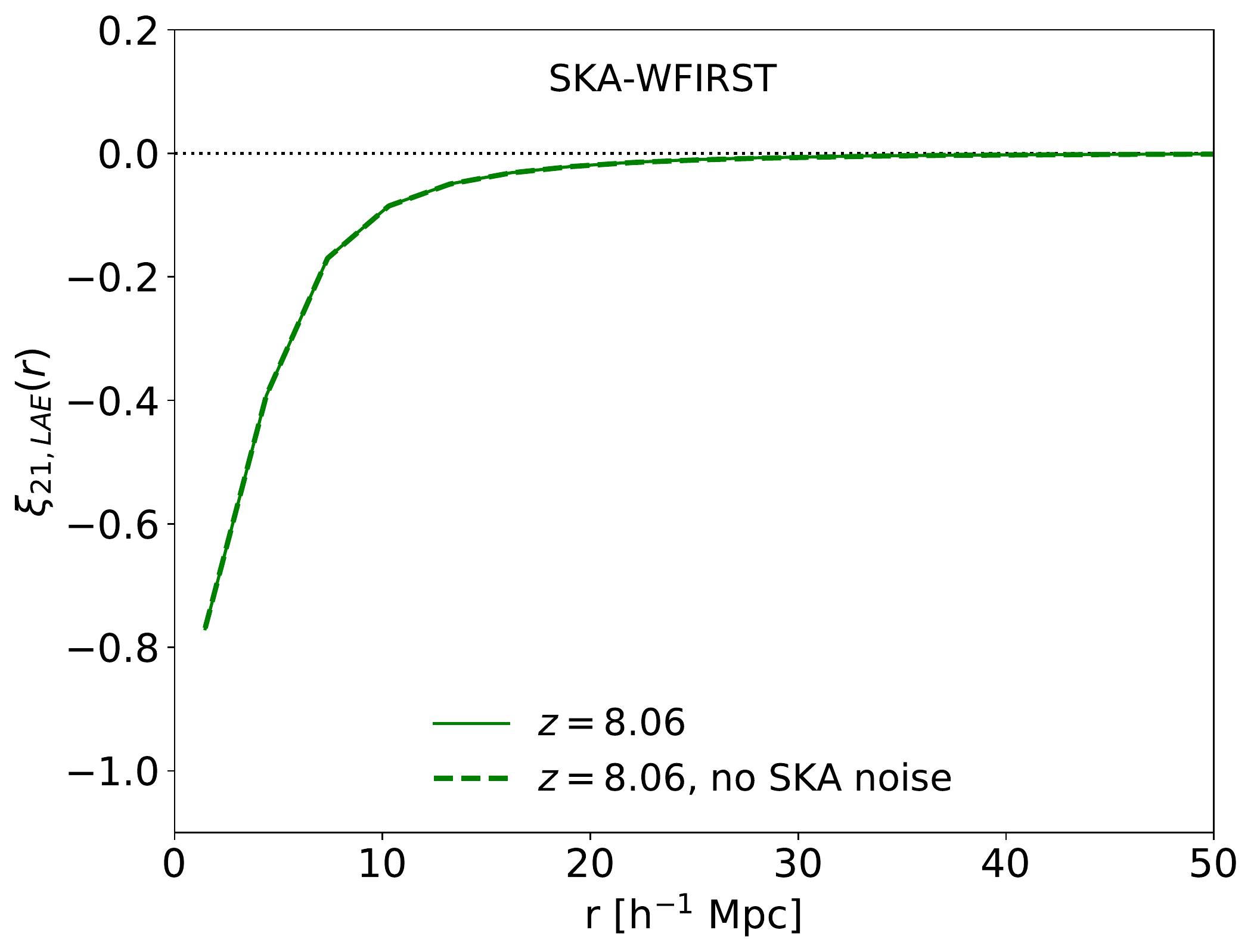}
\caption{3D 21cm-LAE cross-correlation function between SKA and WFIRST with a FoV of 16 deg$^2$ at $z=8.06$. The solid (dashed) line refers to the cross-correlation function with (without) SKA noise. The black dotted line indicates zero correlation.}
\label{SKA:WFIRSTccf}
\end{figure}

Fig. \ref{SKA:WFIRSTccf} shows 21cm-LAE cross-correlation function, which suggests that the average scale of ionized bubbles at the redshift in question is $\sim30 h^{-1}$~Mpc. The addition of the SKA noise hardly makes any difference for the cross-correlation function.

\section{Discussion}
\label{sec:discuss}
\begin{figure}
\centering
\includegraphics[width=84mm]{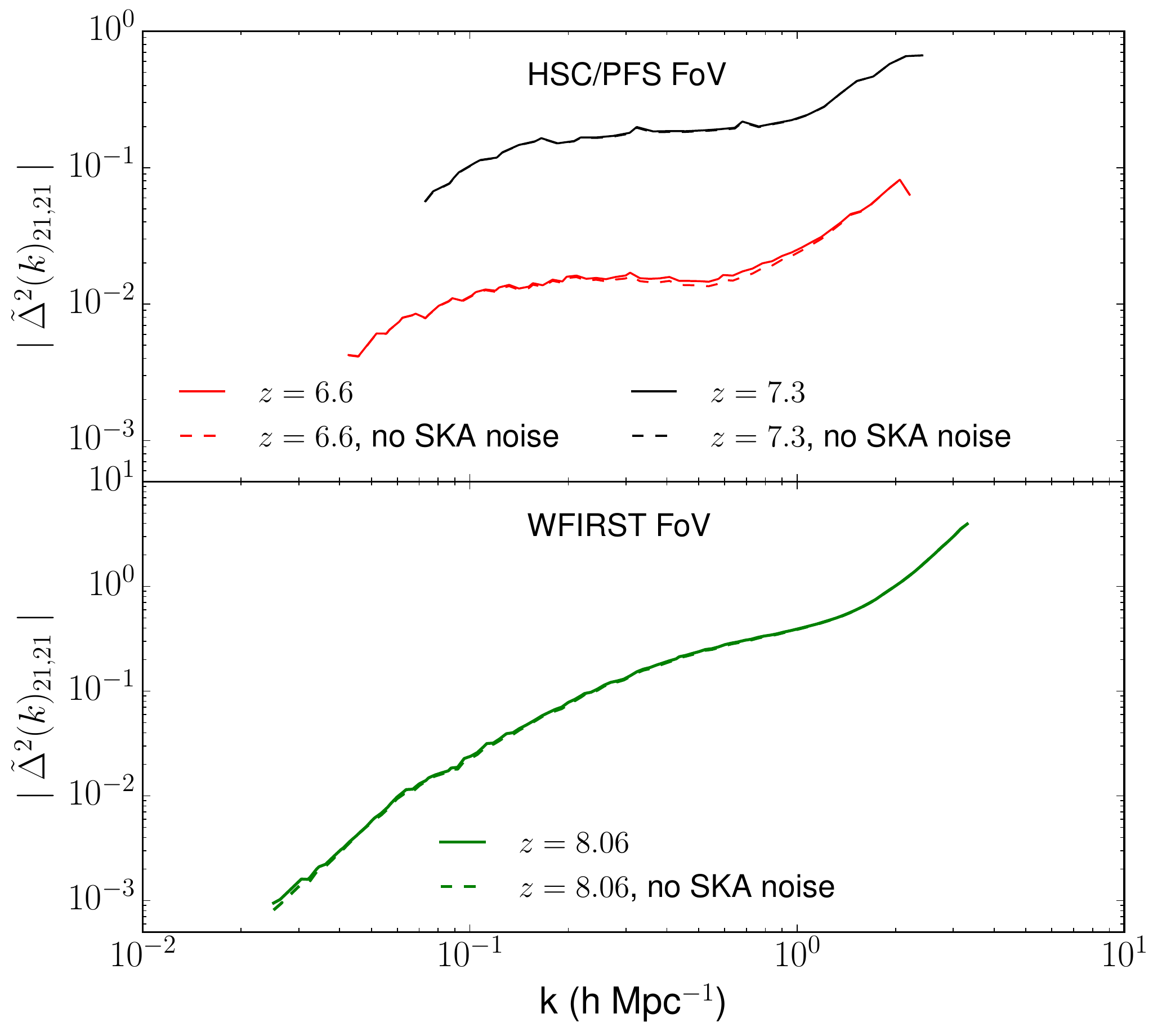}                  
\caption{{\it Top panel}: 2D, unnormalised by $\delta T_{b0}$, 21cm auto-power spectrum at $z=7.3$ (upper set of black lines) and $z=6.6$ (lower set of red lines). The field of view is 1.7 deg$^2$ and 7 deg$^2$ at $z=7.3$ and $z=6.6$, respectively (as HSC and PFS). Solid (dashed) lines refer to the spectrum with (without) SKA noise.
{\it Bottom panel}: 3D, unnormalised by $\delta T_{b0}$, 21cm auto-power spectrum with (solid line) and without (dashed) SKA noise at $z=8.06$. The field of view is 16 deg$^2$ (as WFIRST).
}
\label{SKA:noise}
\end{figure}

The next generation instrument SKA will have much lower noise levels than LOFAR and thus will introduce a significant change when used for the observation of 21cm-LAE cross-correlations.
The 21cm auto-power spectra with and without SKA noise  after 1000 hours of observation in fields of view of 7 deg$^2$ at $z=6.6$ and 1.7 deg$^2$ at $z=7.3$ (i.e. equivalent to the ones of HSC and PFS), and in a field of view of 16 deg$^2$ at $z=8.06$ (i.e. equivalent to the one of WFIRST) are shown in Fig. \ref{SKA:noise}. The difference between results with and without SKA noise is so small it is barely visible, and thus, the main noise contribution to these observations comes from the shot noise due to LAEs observations. The reduction from LOFAR to SKA noise will significantly improve the range of observable scales, from $k<0.1\ h$~Mpc$^{-1}$ to $k<0.3\ h$~Mpc$^{-1}$, as well as the detectable strength of the anti-correlation on the large scales of the cross-correlation coefficient, up to $\sim-0.4$ at $z=7.3$ and $\sim-0.6$ at $z=6.6$. 

\begin{figure}
\centering
\includegraphics[width=84mm]{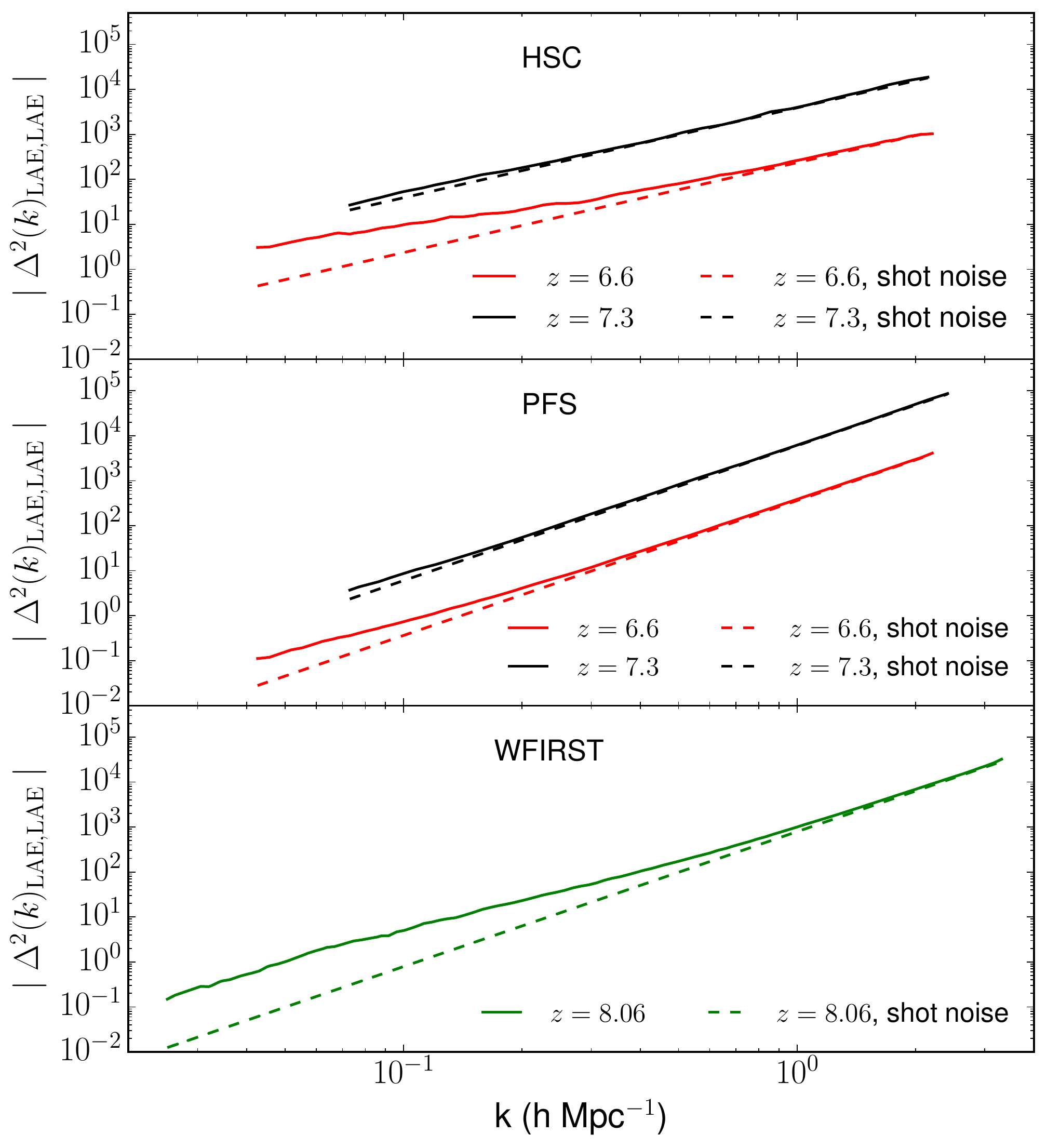}            
\caption{{\it Top panel}: HSC 2D LAE auto-power spectra at $z=7.3$ (black solid line) and $z=6.6$ (red solid), and shot noise power spectra at $z=7.3$ (black dashed) and $z=6.6$ (red dashed). The field of view is 1.7 deg$^2$ and 7 deg$^2$ at $z=7.3$ and $z=6.6$, respectively.
{\it Middle panel}: PFS 3D LAE auto-power spectra at $z=7.3$ (black solid line) and $z=6.6$ (red solid), and shot noise power spectra at $z=7.3$ (black dashed) and $z=6.6$ (red dashed). The field of view is 1.7 deg$^2$ and 7 deg$^2$ at $z=7.3$ and $z=6.6$, respectively.
{\it Bottom panel}: WFIRST 3D LAE auto-power spectrum (green solid line) and shot noise power spectrum (green dashed) at $z=8.06$. The field of view is 16 deg$^2$.
}
\label{shot_noise}
\end{figure}

PFS will introduce significant change compared to HSC by making precise spectroscopic observations of previously detected LAEs, which will enable more precise computations of their positions, thus allowing 3D cross-correlations. However, this important change in LAE information will not result in a significant improvement of the measured cross-correlations, as the same number of LAEs will describe the observed 3D volume poorer than the 2D surface, increasing the shot noise significantly. Thus, the additional information about positions of LAEs and the increase of the shot noise will mostly cancel each other out. 
As shown in Fig. \ref{shot_noise}, the shot noise dominates over the LAE auto-power spectrum more significantly in observations with PFS rather than with HSC. 
At $z=7.3$ with PFS the LAE auto-power spectrum starts to dominate over the shot noise at the same scale as with HSC ($k\sim 0.2\ h$~Mpc$^{-1}$). However, in both cases this dominance is hardly insignificant. 
At $z=6.6$ the LAE auto-power spectrum starts to be relevant on larger scales ($k\sim 0.3\ h$~Mpc$^{-1}$ for PFS and $k\sim 0.7\ h$~Mpc$^{-1}$ for HSC), but the difference with the shot noise in PFS' observations reaches only half an order of magnitude.
Thus, even though in this case there are no projection effects, the shape of the cross-power spectra and the strength of the cross-correlation coefficient do not change significantly due to the simultaneous increase of the shot noise.

In the bottom panel of Fig. \ref{shot_noise} we show the LAE auto-power spectrum and shot noise power spectrum for WFIRST observations at $z=8.06$. The LAE auto-power spectrum dominates over the shot noise power spectrum on scales larger than $k \sim 1\ h$ Mpc$^{-1}$, becoming also an order of magnitude larger. We thus expect that the low noise of both WFIRST and SKA observations should allow a precise measurements on scales larger than $k \sim 1\ h$ Mpc$^{-1}$.

While investigations of the 21~cm-LAEs cross-correlation appeared in the literature give results which are qualitatively similar, a quantitative comparison is somewhat misleading, as the methods and models employed are very different. As discussed in previous literature, the quantitative results depend primarily on the spatial distribution and dimension of HII regions, as well as the population of LAEs. As a consequence, different reionization histories will yield different results, with the choice of the minimum mass of star hosting halos and of how halos are populated with LAEs having a major impact. Also the method used to compute the mock observations and the cross-correlations (e.g. by integrating the power spectra rather than directly) are expected to introduce further differences. 
For these reasons, our results are more in line to those of \cite{Hutter.Dayal.Muller.Trott_2017} and \cite{Kubota_etal_2018}, who employ simulations of reionization more similar to ours, compared e.g. to \cite{SobacchiMesingerGreig2016} who instead use a semi-numeric approach.

\section{Conclusions}

In this paper we have used radiative transfer + N-body simulations to explore the feasibility of measurements of cross-correlations between the 21~cm field observed by the Square Kilometer Array (SKA) and high-$z$ Lyman Alpha Emitters (LAEs) observed in galaxy surveys with the Subaru Hyper Supreme Cam (HSC), Subaru Prime Focus Spectrograph (PFS) and Wide Field Infrared Survey Telescope (WFIRST).
We have investigated both the 21cm-LAE cross-power spectra and cross-correlation functions since the noise affect them differently.
For this study we have closely followed a companion paper based on observations with LOFAR (V2016).

The next generation observations with SKA will have a noise level much lower than those with LOFAR, introducing a significant improvement in the measurement of the cross-correlations. Compared to V2016, we find that an SKA-HSC observation will extend the range of detectable scales from $k<0.1\ h$~Mpc$^{-1}$ to $k<0.3\ h$~Mpc$^{-1}$, as well as increase the strength of the anti-correlation on large scales up to $\sim-0.4$ at $z=7.3$ and $\sim-0.6$ at $z=6.6$, compared to $\sim -0.25$ and $\sim -0.3$ respectively. The cross-correlation functions are much smoother and have a smaller scatter, so that they can be used to investigate scales above $\sim 10\ h^{-1}$~Mpc and $\sim 60\ h^{-1}$~Mpc at $z=7.3$ and 6.6, respectively. 

PFS will introduce  an important improvement to Subaru's measurements by making precise spectroscopic observations of LAEs previously detected with HSC. This will enable a more accurate computation of their positions, thus allowing 3D cross-correlations. However, this additional information will not result in a significant improvement of the measured cross-correlations as the same number of LAEs will describe the observed 3D volume poorer than the 2D surface, increasing the shot noise significantly. Thus, the information gained on the position of LAEs and the increase of the shot noise will mostly cancel each other out.

Unlike Subaru's observations, which will mostly give insight into cross-correlations towards the end of reionization, WFIRST will allow access to also higher redshifts. Specifically, WFIRST is expected to observe spectroscopically about 900 LAEs per square degree and unit redshift in the range $7.5\le z\le 8.5$. As a reference, we have computed the 21cm-LAE cross-correlations at $z=8.06$. As the SKA noise at this redshift is still low and the expected detected LAE density is rather large, the resulting cross-correlations are very similar to the theoretical ones, and shot noise becomes a problem only above $k\sim0.6\ h$~Mpc$^{-1}$. The anti-correlation at large scales reaches a strength of $\sim -0.8$, and the cross-correlation function suggests a typical scale for the ionized regions of $\sim 30\ h^{-1}$~Mpc. In the future, we are planning to investigate how accurately this estimate reproduces the size of the HII regions in the simulations, and its dependence on the method used to quantify it.

In summary, 21cm-LAE cross-correlations are a powerful probe of the EoR and could provide precious information on the progress of reionization and the typical dimension of ionized regions at various redshifts. As 21~cm observations with SKA will be affected by a very small noise, cross-correlations with future galaxy surveys will be feasible in a range of different redshifts. At the same time, WFIRST is expected to observe a large number density of LAEs, reducing the shot noise compared to HSC and PFS. Both effects will result in more precise cross-correlations and increasing observable scales, offering information at various redshifts on e.g. the typical scale of ionized regions.

\section*{Acknowledgements}
The authors would like to thank an anonymous referee for her/his useful comments. The authors acknowledge Paul Shapiro for permission to use the simulations on which this paper was based, described in \cite{Iliev2014}. That work was supported in part by grants and allocations of which Shapiro is the P.I., including US NSF grant AST-1009799, NASA grant NNX11AE09G, NASA/JPL grant RSA Nos 1492788 and 1515294, and supercomputer resources from NSF XSEDE grant TG-AST090005 and the Texas Advanced Computing Center (TACC) at the University of Texas at Austin. Some of the numerical computations were done on the Apollo cluster at The University of Sussex and the Sciama High Performance Compute (HPC) cluster which is supported by the ICG, SEPNet and the University of Portsmouth. Part of the computations were performed on the GPC supercomputer at the SciNet HPC Consortium (courtesy Ue-Li Pen). SciNet is funded by: the Canada Foundation for Innovation under the auspices of Compute Canada; the Government of Ontario; Ontario Research Fund $-$ Research Excellence; and the University of Toronto. The authors thank Kyungjin Ahn for providing the recipe for including subresolution sources in the simulation volume.

\bibliography{literatur}
\bsp

\label{lastpage}
\end{document}